\documentclass[a4paper,10pt,abstracton,openbib,final]{scrartcl}
\usepackage[a4paper,includehead,top=30mm,bottom=30mm,right=30mm,left=30mm]{geometry}
\pdfoutput=1
\usepackage{hyperref}
\hypersetup{pdfauthor={Abali, Barchiesi},pdftitle={Asymptotic homogenization for computing metamaterials parameters}}

\makeatletter
\DeclareOldFontCommand{\bf}{\normalfont\bfseries}{\mathbf}
\DeclareOldFontCommand{\it}{\normalfont\itshape}{\mathit}
\makeatother

\usepackage{amsmath, amsthm, amsxtra}
\usepackage{eucal}
\usepackage{bm, bbm, upgreek, stmaryrd}
\usepackage{booktabs, multirow}
\usepackage[final]{graphicx}
\usepackage{graphics}
\usepackage{mathrsfs}
\usepackage{accents}
\usepackage{amsfonts}
\usepackage{amssymb}
\usepackage{textcomp}
\usepackage{subeqnarray}

\usepackage{mathtools}
\usepackage{nicefrac}
\usepackage{subfig}
\usepackage{xcolor}
\usepackage{url}

\DeclareMathAlphabet{\mathscr}{OT1}{pzc}{m}{it}
\newcommand{\begeq}{\begin{equation}\begin{gathered}}
\newcommand{\eqend}{\end{gathered}\end{equation}}
\newcommand{\begal}{\begin{equation}\begin{aligned}}
\newcommand{\alend}{\end{aligned}\end{equation}}

\renewcommand{\t}[1]{\bm{#1}}

\renewcommand{\d}{\, \mathrm d }

\newcommand{\p}{\partial}
\newcommand{\eps}{\varepsilon}
\newcommand{\del}{\updelta}

\newcommand{\pd}[2]{\frac{\p #1}{\p #2}}

\newcommand{\comma}{\ , \ \ }

\newcommand{\cen}{\overset{\text{c}}}
\newcommand{\mi}{^\text{m}}
\newcommand{\ma}{^\text{M}}
\newcommand{\OO}{\mathcal{O}}

\title{\huge Additive manufacturing introduced substructure and computational determination of metamaterials parameters by means of the asymptotic homogenization}
\dedication{\small Dedicated to Prof. Holm Altenbach on the occasion of his 65$^\text{th}$ birthday}

\author{Bilen Emek Abali$^{1,2}$\thanks{Corresponding author, ORCID: 0000-0002-8735-6071, email: bilenemek@abali.org} \and 
Emilio Barchiesi$^2$ 
\\[0.05in]
\small $^1$Technische Universit\"at Berlin, Institute of Mechanics, MS 2, \\[-0.1in]
\small Einsteinufer 5, 10587 Berlin, Germany
\\
\small $^2$Uppsala University, Division of Applied Mechanics, \\[-0.1in]
\small Department of Materials Science and Engineering, Box 534, SE-751 21 Uppsala, Sweden
\\
\small $^2$International Research Center on Mathematics and Mechanics of Complex Systems, \\[-0.1in]
\small Universit\`a degli Studi dell'Aquila, \\[-0.1in]
\small Via Giovanni Gronchi 18 - Zona industriale di Pile 67100, L'Aquila, Italy
}

\begin{document}
\date{}
\maketitle

\begin{abstract}
Metamaterials exhibit materials response deviation from conventional elasticity. This phenomenon is captured by the generalized elasticity as a result of extending the theory at the expense of introducing additional parameters. These parameters are linked to internal length scales. Describing on a macroscopic level a material possessing a substructure at a microscopic length scale calls for introducing additional constitutive parameters. Therefore, in principle, an asymptotic homogenization is feasible to determine these parameters given an accurate knowledge on the substructure. Especially in additive manufacturing, known under the infill ratio, topology optimization introduces a substructure leading to higher order terms in mechanical response. Hence, weight reduction creates a metamaterial with an accurately known substructure. Herein, we develop a computational scheme using both scales for numerically identifying metamaterials parameters. As a specific example we apply it on a honeycomb substructure and discuss the infill ratio. Such a computational approach is applicable to a wide class substructures and makes use of open-source codes; we make it publicly available for a transparent scientific exchange.
\end{abstract}


\section{Introduction}
Mechanics of metamaterials is gaining an increased interest owing to additive manufacturing technologies allowing us to craft sophisticated structures with different length scales. For weight reduction, material is saved by introducing a substructure. Substructure-related change in materials response is already known \cite{Eringen1964a, mindlin1964, Eringen68}, studied under different assumptions \cite{steinmann1994, eremeyev2012, polizzotto2013a, polizzotto2013b, ivanova2016, abali2018revealing}, and verified experimentally \cite{049,barchiesi2019mechanical,dell2019force,muller2020experimental}. Substructure-related change leads to metamaterials and this phenomenon is explained by theoretical arguments by assuming conventional elasticity in the smaller length scale (microscale) leading to generalized elasticity in the larger length scale (macroscale) \cite{pideri1997second, smyshlyaev2000rigorous, seppecher2011linear, abdoul2018strain, mandadapu2018polar}. 

For constructing theories, different length scales are often incorporated in science. For example, consider the microscale being simply the molecular structure or the lattice structure in a crystalline material conferring anisotropy upon the response at the macroscale \cite{reuss1929berechnung, hashin1962some, shafiro2000anisotropic, lebensohn2004accuracy}. Another prominent structure-related anisotropy occurs in composite materials, where the microscale is composed of fibers and matrix. The alignment of fibers, and how different plies are stacked up, cause the anisotropy as well as values of effective parameters at the macroscale \cite{levin1976determination, willis1977bounds, kushnevsky1998identification, sburlati2018hashin, shekarchizadeh2020experimental}. Porous materials are frequently modeled as a full material with voids as given inclusions at the microscale, we refer to \cite{eshelby1957determination, mori1973average, kanaun1986spherically, hashin1991spherical, nazarenko1996elastic, dormieux2006microporomechanics}. Additive manufacturing is capable of building metamaterials as demonstrated in \cite{kochmann2013homogenized, placidi2016second, turco2017pantographic, solyaev2018numerical, ganzosch20183d, yang2018material}. Also adding texture in 3-D printing introduces a substructure. Especially in metal 3-D printing technologies, the microscale itself is anisotropic \cite{hitzler2018review,wang2019crystallographic,hitzler2019plane}. We emphasize that at the macroscale, in all examples above, the microscale structure is not detectable such that the materials substructure is smeared out that is called homogenization. 

As applied to generalized mechanics, the use of homogenization techniques is challenging \cite{muhlich2012estimation,laudato2020nonlinear}, since generalized mechanics is still evolving \cite{giorgio2019biot,056,muller2020thence}. There exist different homogenization techniques \cite{pideri1997,forest1999estimating,kouznetsova2002multi,pietraszkiewicz2009natural,giorgio2017continuum,dell2016large,maurice2019second}. In generalized mechanics \cite{li2011micromechanics,askes2011gradient}, often a Representative Volume Element (RVE) is exploited as in \cite{willis1981variational,franciosi2018mean}, although the use of an RVE in generalized mechanics is difficult to justify \cite{tran2012micromechanics,rahali2015homogenization}. Yet there exist direct approaches \cite{reda2017dynamical,ganghoffer2019determination} by computational homogenization methods \cite{rahali2016computation,elnady2016computation,rahali2017multiscale,057,063,skrzat2020effective} as well as techniques based on gamma-convergence  \cite{alibert2003truss,alibert2015second}. By means of asymptotic analysis \cite{Bens78, hollister1992comparison, chung2001asymptotic, Temi12} as already applied in \cite{forest2001asymptotic, eremeyev2016effective, ganghoffer2018homogenized, turco2019properties}, we decompose variables into global and local variations \cite{peszynska2007multiscale,pinho2009asymptotic,efendiev2009multiscale} and this separation makes possible to solve the elasticity problem analytically, leading to closed form relations between (known) parameters at the microscale and (sought after) parameters at the macroscale. This approach has been applied in one-dimensional problems for reinforced composites \cite{boutin1996microstructural, barchiesi20181d} and in two-dimensional continuum \cite{bacigalupo2014second, boutin2017linear, placidi2015gedanken, cuomo2014variational} mostly numerically. From extensive studies \cite{peerlings2004computational,li2011establishment,li2013numerical,barboura2018establishment,ameen2018quantitative,porubov2020nonlinear}, we know that this method is adequate for determining metamaterials parameters. We briefly explain the derivation based on \cite{054} and extend the method to the three-dimensional case by providing a numerical procedure by means of the Finite Element Method (FEM). Especially in honeycomb type infill substructure is our interest \cite{tancogne2019stiffness}. The substructure introduces higher order effects as expected and we determine the parameters by a computational homogenization based on the asymptotic analysis. The code uses open-source packages under GNU public license \cite{gnupublic} from the FEniCS project \cite{fenics_www} and we make the code publicly available in \cite{compreal} in order to increase the scientific exchange.

\section{Asymptotic homogenization}

We begin with the assertion that the deformation energy at the microscale is equivalent to the deformation energy at the macroscale,
\begeq
\int_{\Omega} { w\mi} \d V = \int_{\Omega} { w\ma} \d V \ ,\\
\eqend
for the same domain, $\Omega$, occupied by the continuum body. We use the standard continuum mechanics notation with $\d V$ meaning the infinitesimal volume element, expressed in Cartesian coordinates as $\d V=\d x \d y \d z$. There is only one coordinate system used for both scales. We use a material frame, so the location of material particles is denoted by $\t X=(X_1,X_2,X_3)=(x,y,z)$. Furthermore, we use ``m'' and ``M'' denoting microscale and macroscale, respectively. The domains for both scales are equivalent, large enough for allowing homogenization and small enough such that the substructure has a significant effect at the macroscale. We emphasize that a large enough domain---analogously macroscale with a large enough length scale---converges to the classical elasticity approach. 

Since we model an elastic body, the deformation energy depends solely on space derivatives of displacements. At each length scale, there exists one  displacement field, $u\mi_i$, $u\ma_i$. We stress that displacements and their derivatives are different such that the energy density is different in each position. Nevertheless, for the whole body, the total energy is equivalent at both scales. This assertion is the key axiom in nearly all homogenization theories based on the intuition that the energy applied on the body is the same although we observe a different displacement recorded by a 10 MP camera via Digital Image Correlation (DIC) compared to a displacement field captured under a microscope. 

We simplify the analysis by assuming that the system at the microscale is composed of linear elastic material(s) such that the energy is quadratic in displacement gradients given by strains, $\t\eps\mi$, with a known stiffness tensor, $\t C\mi$, as follows:
\begeq
w\mi = \frac{1}{2} C\mi_{ijkl}  \eps\mi_{ij}  \eps\mi_{kl} \comma 
\eps\mi_{ij} = \frac12 ( u\mi_{i,j} + u\mi_{j,i} + u\mi_{k,i} u\mi_{k,j} ) \ ,
\eqend
where the comma denotes a space derivative in $\t X$ and we understand \textsc{Einstein}'s summation convention over repeated indices. For the sake of simplicity, we henceforth use linearized strain measure,
\begeq
\eps\mi_{ij} = \frac12 ( u\mi_{i,j} + u\mi_{j,i} ) \ ,
\eqend
and the usual (minor) symmetries of the stiffness, $C\mi_{ijkl}=C\mi_{jikl}=C\mi_{ijlk}$, we obtain
\begeq
w\mi = \frac{1}{2} C\mi_{ijkl}  u\mi_{i,j}  u\mi_{k,l} \ . 
\eqend
The system at the microscale possesses different materials. For the substructure, for example in an additively manufactured porous structure, we model the structure itself with its stiffness tensor and the voids with a nearly zero stiffness. In other words, the material is heterogeneous at the microscale. At the macroscale, the system is assumed to be homogeneous and to obey materially and geometrically linear strain gradient elasticity modeled by the following deformation energy density:
\begeq
w\ma = \frac12  C\ma_{ijkl}  u\ma_{i,j}  u\ma_{k,l} +  \frac12 D\ma_{ijklmn}  u\ma_{i,jk}  u\ma_{l,mn} + G\ma_{ijklm} u\ma_{i,j} u\ma_{k,lm}\ ,
\eqend
with analogous symmetries $C\ma_{ijkl}=C\ma_{jikl}=C\ma_{ijlk}$ as well as $D\ma_{ijklmn}=D\ma_{ijklmn}=D\ma_{ikjlmn}=D\ma_{lmnijk}$ and $G\ma_{ijklm}=G\ma_{jiklm}=G\ma_{ijkml}$. We stress that $\t G\ma=0$ if the macroscale is of centro-symmetric substructure and $\t D\ma=0$ leads to conventional elasticity with substructure related anisotropy without higher order (strain gradient) terms. 

First, we introduce a so-called geometric center:
\begeq \label{center_def}
\cen{\t X}  = \frac1V \int_{\Omega} \t{X} \d V \ ,
\eqend
and, assuming enough continuity, approximate the macroscale displacement by the \textsc{Taylor} expansion around the value at the geometric center by truncating after quadratic terms. The choice of quadratic terms is justified by the nonlocality of the theory, in other words, we aim for the strain gradient theory incorporating second derivatives. All higher terms than the second derivative will be neglected. The expansion of displacement gradients reads
\begeq\label{expansion1}
{u\ma _i}(\t{X}) = {u\ma _i}\Big|_{\cen{\t X}} 
+  u\ma_{i,j} \Big|_{\cen{\t X}} (X_j-\cen X_j) 
+ \frac{1}{2}  u\ma_{i,jk} \Big|_{\cen{\t X}} (X_j-\cen X_j) (X_k-\cen X_k)  \ .
\eqend
Since ${u\ma _i}\Big|_{\cen{\t X}}$ is a vector evaluated at $\cen{\t X}$, its gradient vanishes leading to
\begal \label{expansion2}
u\ma_{i,l}(\t{X}) &= u\ma_{i,j} \Big|_{\cen{\t X}} \delta_{jl}
+ \frac{1}{2}  u\ma_{i,jk} \Big|_{\cen{\t X}} ( \delta_{jl} (X_k-\cen X_k) + (X_j-\cen X_j) \delta_{kl} ) 
\ , \\
&= u\ma_{i,l} \Big|_{\cen{\t X}} +  u\ma_{i,lk} \Big|_{\cen{\t X}} (X_k-\cen X_k)
\ , \\
u\ma_{i,lm}(\t{X}) &= u\ma_{i,lk} \Big|_{\cen{\t X}} \delta_{km} = u\ma_{i,lm} \Big|_{\cen{\t X}} \ .
\alend
Second, we introduce spatial averaging for displacement gradients by using the latter expansions and the fact that terms evaluated at $\cen{\t X}$ are constant within the domain
\begal \label{displacement average}
	\langle {u}_{i,j}\ma \rangle &=\frac{1}{V}\int_{\Omega} u\ma _{i,j} \d V= u\ma_{i,j}\Big|_{\cen{\t X}} + u\ma_{i,jk}\Big|_{\cen{\t X}} \bar I_k 
	\ , \\
	\langle {u}_{i,jk}\ma \rangle &=\frac{1}{V}\int_{\Omega} u\ma _{i,jk}\d V= u\ma_{i,jk} \Big|_{\cen{\t X}} \ ,
\alend
with 
\begeq
\bar I_k = \frac1V \int_{\Omega} (X_k-\cen X_k) \d V = \frac1V \int_{\Omega} X_k \d V - \frac1V \int_\Omega \cen X_k \d V = 0 \ ,
\eqend
since integration is additive and we have inserted Eq.\,\eqref{center_def}. Thus, we obtain
\begeq \label{displacement average approx}
	\langle {u}_{i,j}\ma \rangle =  u\ma_{i,j}\Big|_{\cen{\t X}} \comma
	\langle {u}_{i,jk}\ma \rangle = u\ma_{i,jk} \Big|_{\cen{\t X}} \ .
\eqend
Third, we use the spatial averaged values in the expansions \eqref{expansion1} and \eqref{expansion2}  
\begal \label{macro.relations}
	u_i\ma(\t X) &= u\ma _i\Big|_{\cen{\t X}} + \langle {u}_{i,j}\ma \rangle (X_j-\overset{\text{c}}X_j) +  \frac{1}{2} \langle {u}_{i,jk}\ma \rangle (X_j-\overset{\text{c}}X_j) (X_k-\overset{\text{c}}X_k)  \ , \\
	u\ma _{i,j}(\t X) &= \langle{u}_{i,j}\ma \rangle + \langle{u}_{i,jk}\ma\rangle (X_k-\overset{\text{c}}X_k) \ , \\
	u\ma_{i,jk}(\t X) &= \langle {u}_{i,jk}\ma  \rangle \ .
\alend
Obviously, we circumvent using any spatial averaging techniques \cite{whitaker1967diffusion,slattery1967flow,gray1977theorems}. Finally, we insert the latter into the energy definition and take out spatial averaged terms out of the integral
\begal \label{macro.energy.definition}
\int_\Omega w\ma \d V =&
\int_\Omega \bigg( \frac12 C\ma_{ijlm} u\ma_{i,j} u\ma_{l,m} + \frac12 D\ma_{ijklmn} u\ma_{i,jk} u\ma_{l,mn} + G\ma_{ijklmn} u\ma_{i,j} u\ma_{k,lm}  \bigg) \d V
 \\ 
=& \frac12 C\ma_{ijlm}  \int_{\Omega} u\ma_{i,j} u\ma_{l,m} \d V
+ \frac12 D\ma_{ijklmn} \int_{\Omega} u\ma_{i,jk} u\ma_{l,mn} \d V
+ G\ma_{ijklm} \int_{\Omega} u\ma_{i,j} u\ma_{k,lm} \d V 
 \\
=& \frac12 C\ma_{ijlm} \int_{\Omega} \Big( \langle{u}_{i,j}\ma \rangle +\langle {u}_{i,jk}\ma \rangle (X_k-\cen X_k) \Big) 
 \Big(\langle{u}_{l,m}\ma \rangle + \langle{u}_{l,mn}\ma\rangle (X_n-\cen X_n) \Big) \d V 
+ \\
&+ \frac12 D\ma_{ijklmn} \int_{\Omega} \langle{u}_{i,jk}\ma\rangle \langle{u}_{l,mn}\ma\rangle  \d V 
+ G\ma_{ijlmn} \int_{\Omega} \Big( \langle{u}_{i,j}\ma \rangle +\langle {u}_{i,jk}\ma \rangle (X_k-\cen X_k) \Big) \langle{u}_{l,mn}\ma\rangle  \d V 
 \\
=&  \frac{V}{2} \bigg(  C\ma_{ijlm} \langle{u}_{i,j}\ma\rangle \langle{u}_{l,m}\ma\rangle 
+  \big( C\ma_{ijlm} \bar{I}_{kn} +  D\ma_{ijklmn} + 2 G\ma_{ijlmn} (X_k-\cen X_k) \big) \langle{u}_{i,jk}\ma\rangle \langle{u}_{l,mn}\ma\rangle 
+ \\
&+ 2 G\ma_{ijlmn} \langle{u}_{i,j}\ma \rangle  \langle u\ma_{l,mn}\rangle \bigg) \ , \\
\alend
by using 
\begeq
\bar{I}_{kn} = \frac1V \int_{\Omega} (X_k-\cen X_k)(X_n-\cen X_n) \d V \ .
\eqend
By following the asymptotic homogenization method, we use a so-called homothetic ratio, $\epsilon$, for a separation of length scales and introduce the local coordinates,
\begin{equation} \label{link_coord}
y_j= \frac1\epsilon ( X_j - \cen X_j ) \ .
\end{equation}
Therefore, the macroscale relations in Eq.\,\eqref{macro.relations} become
\begal \label{macro.relations2}
	u_i\ma(\t X) &= u\ma _i\Big|_{\cen{\t X}} + \epsilon y_j \langle {u}_{i,j}\ma \rangle +  \frac{1}{2} \epsilon^2 y_j y_k \langle {u}_{i,jk}\ma \rangle  \ , \\
	u\ma _{i,j}(\t X) &= \langle{u}_{i,j}\ma \rangle + \epsilon y_k \langle{u}_{i,jk}\ma\rangle  \ , \\
	u\ma_{i,jk}(\t X) &= \langle {u}_{i,jk}\ma  \rangle \ .
\alend
With the assumption that the displacement field is a smooth function at the macroscale and $\t y$-periodic in local coordinates, the mean local fluctuations vanish within the chosen domain, $\Omega$. In other words, the effective property at the macroscale is constant representing the ``oscillatory'' property at the microscale. The difference between the effective (macroscale) and oscillatory (microscale) property is the fluctuation to vanish. In this regard, we decompose the microscale displacement
\begin{equation} \label{displacement_function}
\t{u \mi}(\t X) = \overset{0} {\t u}( \t X, \t y) + \epsilon \overset{1}{\t u}(\t X,\t y) + \epsilon^2 \overset{2} {\t u}(\t X, \t y) + \OO(\epsilon^3) \ ,
\end{equation}
where $\overset{n} {\t u}( \t X, \t y)$ ($n$ = 0, 1, 2) are $\t y$-periodic. In other words, the chosen domain, $\Omega$, acts as a Representative Volume Element (RVE) within that we seek the effective property. 

We use the well-known least action principle for solving the displacement by starting off with the \textsc{Lagrange} function, $\rho f_i u\mi_i - w\mi$, where the gravitational specific (per mass) force, $f_i$, and the mass density, $\rho$, are given. For finding the variation of the action functional by the arbitrary test functions, $\del \t u$, we perform an integration by part where the domain boundaries, $\p\Omega$, are identical to those from neighboring RVEs. Since the normal vectors, $\t n$, of neighboring surfaces, $\d A$, are opposite, all boundaries vanish
\begal\label{equilibrium}
0 =& \del \int_\Omega \big( \rho f_i u\mi_i - w\mi \big) \d V  \ , \\
0 =& \int_\Omega \big( \rho f_i \del u\mi_i - C\mi_{ijkl}  u\mi_{k,l} \del u\mi_{i,j} \big) \d V  \ , \\
0 =& \int_\Omega \Big( \rho f_i  + \big( C\mi_{ijkl}  u\mi_{k,l} \big)_{,j} \Big) \del u\mi_i \d V - \int_{\p\Omega} C\mi_{ijkl}  u\mi_{k,l} n_j \del u\mi_i \d A  \ , \\
0 =& \rho f_i + \big( C\mi_{ijkl}  u\mi_{k,l} \big)_{,j}  \ .
\alend
Derivative of the microscale displacement from Eq.\,\eqref{displacement_function} after inserting Eq.\,\eqref{link_coord} reads
\begal
u\mi_{i,j} 
=& \Big( \overset{0} u_i( \t X, \t y) + \epsilon \overset{1}u_i (\t X,\t y) + \epsilon^2 \overset{2}u_i(\t X, \t y) + \OO(\epsilon^3) \Big)_{,j}
 \\
=& \overset{0} u_{i,j} + \epsilon \overset{1}u_{i,j} + \epsilon^2 \overset{2}u_{i,j}  
+ \frac{\delta_{kj}}{\epsilon} \pd{}{y_k} \Big( \overset{0} u_i + \epsilon \overset{1}u_i + \epsilon^2 \overset{2}u_i  \Big) + \OO(\epsilon^3)
 \\
=& \overset{0} u_{i,j} + \pd{\overset{0} u_i}{y_j} \frac1\epsilon 
+ \epsilon \overset{1}u_{i,j} + \pd{\overset{1}u_i}{y_j}
+ \epsilon^2 \overset{2}u_{i,j} + \epsilon \pd{\overset{2}u_i}{y_j}  + \OO(\epsilon^3) \ .
\alend
Inserting the latter in Eq.\,\eqref{equilibrium} and once more using the chain rule in combination with Eq.\,\eqref{link_coord}, we obtain
\begeq
\rho f_i +  \Bigg(
C\mi_{ijkl} \bigg( \overset{0} u_{k,l} + \pd{\overset{0} u_k}{y_l} \frac1\epsilon 
+ \epsilon \overset{1}u_{k,l} + \pd{\overset{1}u_k}{y_l}
+ \epsilon^2 \overset{2}u_{k,l} + \epsilon \pd{\overset{2}u_k}{y_l} \bigg)
\Bigg)_{,j}
+ \\
+
\frac1\epsilon \pd{}{y_j}
\Bigg(
C\mi_{ijkl} \bigg( \overset{0} u_{k,l} + \pd{\overset{0} u_k}{y_l} \frac1\epsilon 
+ \epsilon \overset{1}u_{k,l} + \pd{\overset{1}u_k}{y_l}
+ \epsilon^2 \overset{2}u_{k,l} + \epsilon \pd{\overset{2}u_k}{y_l} \bigg)
\Bigg) =0
\eqend
where separation of coefficients multiplied by the same order in $\epsilon$ and setting every term zero---since $\epsilon$ and $\epsilon^2$ terms are independent---results in
\begeq\label{additional.conditions}
\frac1{\epsilon^2} \pd{}{y_j} \Big( C_{ijkl}\mi \pd{ \overset{0}u_k}{y_l} \Big) = 0 \ ,
\\
\frac1\epsilon \Bigg(
\bigg( C\mi_{ijkl} \pd{\overset{0}u_k}{y_l} \bigg)_{,j} 
+  \pd{}{y_j} \big( C_{ijkl}\mi   \overset{0}u_{k,l} \big)  
+ \pd{}{y_j} \Big( C\mi_{ijkl}   \pd{\overset{1}u_k}{y_l} \Big)
\Bigg)   
= 0 \ ,
\\
\rho f_i + \big( C\mi_{ijkl}   \overset{0}u_{k,l} \big)_{,j} 
+  \Big( C\mi_{ijkl}   \pd{\overset{1}u_k}{y_l} \Big)_{,j}
+  \pd{}{y_j} \big( C_{ijkl}\mi  \overset{1}u_{k,l} \big)
+ \pd{}{y_j} \Big( C_{ijkl}\mi  \pd{\overset{2}u_k}{y_l} \Big)   
= 0 \ ,
\\
\epsilon \Bigg(
\big( C\mi_{ijkl}   \overset{1}u_{k,l} \big)_{,j} 
+ \Big( C_{ijkl}\mi \pd{ \overset{2}u_k}{y_l} \Big)_{,j}
+  \pd{}{y_j} \big( C_{ijkl}\mi  \overset{2}u_{k,l} \big)
\Bigg) =0 \ ,
\\
\epsilon^2 \big( C\mi_{ijkl}  \overset{2}u_{k,l} \big)_{,j} = 0 \ .
\eqend
Since $C\mi_{ijkl}$ depends on $\t y$, for example consider two distinct materials at the microscale, from the first relation, we immediately conclude that $\overset{0}u_i=\overset{0}u_i(\t X)$. By using this dependency, we introduce the multiplicative decomposition
\begeq
\overset{1}u_i = \overset{0}u_{a,b}(\t X) \varphi_{abi}(\t y) \comma
\overset{2}u_i = \overset{0}u_{a,bc}(\t X) \psi_{abci}(\t y) \ ,
\eqend
with the unknown tensors $\varphi_{abc}$ and $\psi_{abcd}$. The latter decomposition is a general procedure in tensor calculus and the unknown tensors, $\t\varphi$, $\t\psi$, have no underlying assumptions. As a consequence, for $\t u\mi$, we have the following expression:
\begeq\label{micro.expansion}
u\mi_i = \overset{0}u_i(\t X) + \epsilon \overset{0}u_{a,b}(\t X) \varphi_{abi}(\t y)  + \epsilon^2 \overset{0}u_{a,bc}(\t X) \psi_{abci}(\t y) + \OO(\epsilon^3) \ ,
\eqend
with the first term---the sole term depending only on $\t X$, all the other terms depend on $\t y$ as well---corresponding to the macroscale displacement, 
\begeq\label{macro.zero.relation}
\t u\ma = \overset{0}{\t u}(\t X) \ .
\eqend 
By using Eq.\,\eqref{macro.zero.relation} in Eq.\,\eqref{micro.expansion}, we obtain the displacement gradient,
\begal 
u\mi_{i,j} &= 
\Big( u\ma_i + 
	\epsilon u\ma_{a,b} \varphi_{abi} + 
	\epsilon^2 u\ma_{a,bc} \psi_{abci} \Big)_{,j} + \OO(\epsilon^3) 
\\
	&= u_{i,j}\ma 
	+ \frac{\p \varphi_{abi}}{\p y_j}  u\ma_{a,b}
	+ \epsilon \varphi_{abi} u\ma_{a,bj}
	+ \epsilon \frac{\p \psi_{abci}}{\p y_j} u\ma_{a,bc} 
	+ \epsilon^2 \psi_{abci} u\ma_{a,bcj} + \OO(\epsilon^3)
\\
	&= \underbrace{\Big(\delta_{ia} \delta_{jb} + \frac{\p \varphi_{abi}}{\p y_j} \Big)}_{L_{abij}} u\ma_{a,b} 
	+ \epsilon u\ma_{a,bc} \underbrace{\Big( \varphi_{abi} \delta_{jc} + \frac{\p \psi_{abci} }{\p y_j} \Big)}_{N_{abcij}} 
	+ \epsilon^2 \psi_{abci} u\ma_{a,bcj} + \OO(\epsilon^3) \ ,
\alend
and, after inserting Eq.\,\eqref{macro.relations2}, we acquire
\begeq
u\mi_{i,j} = L_{abij} \langle u\ma_{a,b} \rangle + \epsilon \langle u\ma_{a,bc} \rangle y_c L_{abij}  + \epsilon  \langle u\ma_{a,bc} \rangle N_{abcij} \ ,
\eqend
since we incorporate up to the second gradients in Eq.\,\eqref{expansion1}. By using $M_{abcij} = y_c L_{abij} + N_{abcij}$ we calculate the energy at the microscale
\begal
\int_{\Omega} w\mi \d V 
=&
\frac12 \int_{\Omega^P} \Big( C_{ijkl}\mi L_{abij} L_{cdkl} \langle{u}_{a,b}\ma\rangle \langle{u}_{c,d}\ma \rangle
	+2 \epsilon C_{ijkl}\mi L_{abij} M_{cdekl} \langle{u}_{a,b}\ma\rangle \langle{u}_{c,de}\ma\rangle 
+ \\
&	+ \epsilon^2 C_{ijkl}\mi M_{abcij} M_{defkl} \langle{u}_{a,bc}\ma\rangle \langle{u}_{d,ef}\ma\rangle 
\Big) \d V \\
=& \frac{V}{2}\Big( \bar{C}_{abcd} \langle{u}_{a,b}\ma\rangle \langle{u}_{c,d}\ma\rangle 
	+ 2\bar{G}_{abcde}\langle{u}_{a,b}\ma\rangle \langle{u}_{c,de}\ma\rangle 
	+ \bar{D}_{abcdef}\langle{u}_{a,bc}\ma\rangle \langle{u}_{d,ef}\ma\rangle 
	\Big)  \ .
\alend
with
\begeq \label{relation_to_epsilon}
\bar{C}_{abcd} = \frac{1}{V} \int_{\Omega}  C_{ijkl}\mi L_{abij} L_{cdkl} \d V \ , \\
\bar{G}_{abcde} = \frac{\epsilon}{V} \int_{\Omega}  C_{ijkl}\mi L_{abij} M_{cdekl} \d V \ , \\
\bar{D}_{abcdef} = \frac{\epsilon^2}{V} \int_{\Omega} C_{ijkl}\mi M_{abcij} M_{defkl} \d V \ .
\eqend
Immediately we observe by comparing with Eq.~\eqref{macro.energy.definition}, 
\begeq \label{relation_of_stiffness}
	C\ma_{ijlm} = \bar{C}_{ijlm} \ , \\
	G\ma_{ijlmn} = \bar{G}_{abcde} \ , \\
	C\ma_{ijlm} \bar{I}_{kn} + D\ma_{ijklmn} + 2 \epsilon y_k G\ma_{ijlmn}= \bar{D}_{ijklmn}  \ ,
\eqend
where 
\begeq
\bar{I}_{kn} = \int_{\Omega^P} (X_k-\overset{\text{c}}X_k)(X_n-\overset{\text{c}}X_n) \d V =\epsilon^2 \int_{\Omega^P}y_k y_n \d V \ .
\eqend
Therefore, $\t C\ma$, $\t D\ma$, $\t G\ma$ are determined once $\t\varphi$ and $\t\psi$ are calculated by using the substructure. For these variables, we will obtain corresponding field equations in the following.

By inserting Eq.\,\eqref{micro.expansion} in Eq.\,\eqref{additional.conditions}$_{2}$ and using $\overset{0}{\t u}=\overset{0}{\t u}(\t X)$, we obtain
\begal\label{condition.final.1}
\pd{}{y_j} \big( C\mi_{ijkl}   \overset{0}u_{k,l} \big)  
+ \pd{}{y_j} \Big( C\mi_{ijkl}   \pd{\overset{0}u_{a,b}\varphi_{abk}}{y_l} \Big) 
=& 0 \ , \\
\pd{C\mi_{ijkl}}{y_j} \delta_{ak}\delta_{bl} \overset{0}u_{a,b}   
+ \pd{}{y_j} \Big( C\mi_{ijkl}   \pd{\varphi_{abk}}{y_l} \Big) \overset{0}u_{a,b} 
=& 0 \ , \\
\pd{}{y_j} \Bigg( C\mi_{ijkl} \underbrace{\Big( \delta_{ak}\delta_{bl} + \pd{\varphi_{abk}}{y_l} \Big) }_{L_{abkl}} \Bigg) =& 0 \ .
\alend
Analogously, by exploiting Eq.\,\eqref{additional.conditions}$_{3}$ and inserting the latter, we acquire
\begal\label{additionalconst3}
\rho f_i + \big( C\mi_{ijkl}   \overset{0}u_{k,l} \big)_{,j} 
+  \Big( C\mi_{ijkl}   \pd{\overset{0}u_{a,b} \varphi_{abk}}{y_l} \Big)_{,j}
+  \pd{}{y_j} \big( C\mi_{ijkl}  \overset{0}u_{a,bl}\varphi_{abk} \big)
+ \pd{}{y_j} \Big( C\mi_{ijkl}  \pd{\overset{0}u_{a,bc}\psi_{abck}}{y_l} \Big)   
=& 0 
\ , \\
\rho f_i + C\mi_{ijkl}   \overset{0}u_{k,lj} 
+  C\mi_{ijkl} \overset{0}u_{a,bj} \pd{\varphi_{abk}}{y_l} 
+  \pd{}{y_j} \big( C\mi_{ijkl}  \varphi_{abk} \big) \overset{0}u_{a,bl}
+ \pd{}{y_j} \Big( C\mi_{ijkl}  \pd{\psi_{abck}}{y_l} \Big) \overset{0}u_{a,bc}   
=& 0 
\ , \\
\rho f_i + C\mi_{ickl} \overset{0}u_{a,bc} \underbrace{\Bigg(
  \delta_{ak} \delta_{bl} 
+   \pd{\varphi_{abk}}{y_l} 
\Bigg)}_{L_{abkl}}
+ \overset{0}u_{a,bc} \pd{}{y_j} \Bigg(
C\mi_{ijkl} 
\underbrace{ \Big(  \varphi_{abk} \delta_{cl}
+  \pd{\psi_{abck}}{y_l} \Big) }_{N_{abckl}}
\Bigg)    
=& 0 
\alend
Equations\,\eqref{additional.conditions}$_{4,5}$ are identically fulfilled 
\begal
\big( C\mi_{ijkl}   \overset{1}u_{k,l} \big)_{,j} 
+ \Big( C\mi_{ijkl} \pd{ \overset{2}u_k}{y_l} \Big)_{,j}
+ \pd{}{y_j} \Big( C\mi_{ijkl} \overset{2}u_{k,l} \Big)
=& 0 
\ , \\
C\mi_{ijkl}   \overset{0}u_{a,blj} \varphi_{abk}
+ C\mi_{ijkl} \pd{ \overset{0}u_{a,bcj} \psi_{abck}}{y_l} 
+ \pd{}{y_j} \Big( C\mi_{ijkl} \overset{0}u_{a,bcl} \psi_{abck} \Big)
=& 0 
\ , \\
C\mi_{ijkl} \overset{2}u_{k,lj} =& 0 \ , \\
C\mi_{ijkl} \overset{0}u_{a,bclj} \psi_{abck} =& 0 \ , 
\alend 
since we incorporate only up to the second derivative in Eq.\,\eqref{expansion1}.

In the case of the macroscale, with the least action principle by means of the \textsc{Lagrange} function, $\rho f_i u\ma_i - w\ma$, we obtain after using integration by parts twice and letting the domain boundaries vanish
\begal
0 =& \del \int_\Omega \big( \rho f_i u\ma_i - w\ma \big) \d V  \ , \\
0 =& \int_\Omega \Big( \rho f_i \del u\ma_i - C\ma_{ijkl}  u\ma_{k,l} \del u\ma_{i,j} - D\ma_{ijklmn} u\ma_{l,mn} \del u\ma_{i,jk} 
- G\ma_{ijklm} \del u\ma_{i,j} u\ma_{k,lm} - G\ma_{ijklm} u\ma_{i,j} \del u\ma_{k,lm} \Big) \d V  \ , \\
0 =& \rho f_i + C\ma_{ijkl}  u\ma_{k,lj}  - D\ma_{ijklmn} u\ma_{l,mnjk} + G\ma_{ijklm} u\ma_{k,lmj} - G\ma_{kjilm} u\ma_{k,jlm}  \ , \\
0 =& \rho f_i + C\ma_{ijkl}  u\ma_{k,lj}  \ ,
\alend
since the stiffness tensors are constant at the macroscale, as well as we incorporate only up to the second derivative in Eq.\,\eqref{expansion1}. By using this relation in Eq.\,\eqref{additionalconst3}, we get
\begal\label{condition.final.2}
-C\ma_{icab}  u\ma_{a,bc}
+ C\mi_{ickl} \overset{0}u_{a,bc} L_{abkl}
+ \overset{0}u_{a,bc} \pd{}{y_j} \Bigg(
C\mi_{ijkl} N_{abckl}
\Bigg)    
=& 0 \ , \\
-C\ma_{icab} 
+ C\mi_{ickl} L_{abkl}
+ \pd{}{y_j} \Big(
C\mi_{ijkl} N_{abckl}
\Big)    
=& 0 \ .
\alend
By solving Eq.\eqref{condition.final.1} and Eq.\,\eqref{condition.final.2}$_2$, we calculate $\t\varphi$ and $\t\psi$.

\section{Method of solution}

We sum up the methodology proposed herein. Consider a metamaterial with a given substructure at the microscale, $\t y$. Modeling the substructure with the given $\t C\mi$ by means of the finite element method leads to a numerical solution of $\t\varphi$ by satisfying Eq.\eqref{condition.final.1}:
\begeq\label{solve1}
\pd{}{y_j} \Bigg( C\mi_{ijkl}  L_{abkl}  \Bigg) = 0 \comma
L_{abkl} = \delta_{ak}\delta_{bl} + \pd{\varphi_{abk}}{y_l}
 \ .
\eqend
By using the solution, from Eqs.\,\eqref{relation_to_epsilon}, \eqref{relation_of_stiffness}, we determine
\begeq\label{solve2}
C\ma_{abcd} = \bar{C}_{abcd} = \frac{1}{V} \int_{\Omega}  C_{ijkl}\mi L_{abij} L_{cdkl} \d V \ .
\eqend
The macroscale stiffness tensor, $\t C\ma$, is used in Eq.\eqref{condition.final.2}$_2$ in order to acquire $\t\psi$ by fulfilling
\begeq\label{solve3}
-C\ma_{icab} 
+ C\mi_{ickl} L_{abkl}
+ \pd{}{y_j} \Big(
C\mi_{ijkl} N_{abckl}
\Big)    
= 0 \comma
N_{abckl} = \varphi_{abk} \delta_{cl}
+  \pd{\psi_{abck}}{y_l} \ .
\eqend
With this solution, we construct
\begeq\label{solve4}
M_{abcij} = y_c L_{abij} + N_{abcij} \comma
\bar{I}_{kn} =\epsilon^2 \int_{\Omega^P}y_k y_n \d V \ .
\eqend
and determine
\begeq\label{solve5}
G\ma_{abcde} = \bar{G}_{abcde} = \frac{\epsilon}{V} \int_{\Omega}  C_{ijkl}\mi L_{abij} M_{cdekl} \d V \ , \\
\bar{D}_{abcdef} = \frac{\epsilon^2}{V} \int_{\Omega} C_{ijkl}\mi M_{abcij} M_{defkl} \d V \ , \\
D\ma_{ijklmn} = \bar{D}_{ijklmn} - C\ma_{ijlm} \bar{I}_{kn} - 2 \epsilon y_k G\ma_{ijlmn} \ .
\eqend
The outcome is determining the components of $\t C\ma$ tensor of rank four, $\t G\ma$ tensor of rank five, and $\t D\ma$ tensor of rank six.

In particular, for the numerical solution of Eq.\,\eqref{solve1} as well as Eq.\,\eqref{solve3}, we follow the standard procedure of the finite element method \cite{zohdi2018finite} and utilize a finite dimensional \textsc{Hilbert}ian \textsc{Sobolev} space for trial functions. The same space is used for the test functions as well, called the \textsc{Galerkin} procedure. The triangulation of the structure in $\t y$ is established by using tetrahedrons, and we solve the discrete problem by minimizing the weak form. In order to get the weak forms, Eqs.\,\eqref{solve1},\eqref{solve3} are multiplied by arbitrary test functions of their ranks for reducing to a scalar integrated over the volume of the structure, $\Omega$. For fulfilling the $\t y$ periodicity, all boundaries are modeled as periodic boundaries by tying the nodes on corresponding surfaces. In other words, for a cube from left to right along $X_1$-axis, each node, say, on the left surface has to have the same displacement as its counterpart with the same $X_2$, $X_3$ coordinates on the right surface. Hence, technically, all boundaries are of \textsc{Dirichlet} type and the test functions vanish on all boundaries, for an alternative approach of weak periodicity, we refer to \cite{larsson2011computational}. We use herein a strong coupling with the same mesh on corresponding boundaries, since we use the RVE only at the level of parameter determination. 

All the implementation is carried out in the FEniCS platform, we refer to \cite{027} for an introduction with examples. The weak form is obtained after integrating by parts, we stress that the periodic boundary condition causes that boundary integrals vanish. Moreover, we omit distinguishing between the functions and their discrete representations, since they never occur in the same equation. In order to calculate $\t\varphi$ and $\t\psi$, by utilizing Eq.\eqref{condition.final.1} and Eq.\,\eqref{condition.final.2}$_2$, we obtain the following weak forms:
\begeq\label{solve6}
\int_\Omega  C\mi_{ijkl}  L_{abkl}  \pd{\del \varphi_{abi}}{y_j} \d V = 0 \comma \\
\int_\Omega \bigg( -C\ma_{icab} \del\psi_{abci} 
+ C\mi_{ickl} L_{abkl} \del\psi_{abci} 
-  \Big(
C\mi_{ijkl} N_{abckl}
\Big) \pd{\del\psi_{abci} }{y_j} \bigg) \d V
= 0 \comma
\eqend
are solved separately by setting $a$,$b$,$c$ indices. This fact is of importance so we write out explicitly, how it is meant to do. Because of the minor symmetry, $C\ma_{ijkl}=C\ma_{ijlk}$, we know that $L_{abkl}=L_{bakl}$ and $\varphi_{abi}=\varphi_{bai}$ such that we solve six weak forms
\begeq
\int_\Omega  C\mi_{ijkl}  L_{11kl}  \pd{\del \varphi_{11i}}{y_j} \d V = 0 \comma 
\int_\Omega  C\mi_{ijkl}  L_{22kl}  \pd{\del \varphi_{22i}}{y_j} \d V = 0 \comma 
\int_\Omega  C\mi_{ijkl}  L_{33kl}  \pd{\del \varphi_{33i}}{y_j} \d V = 0 \comma \\
\int_\Omega  C\mi_{ijkl}  L_{23kl}  \pd{\del \varphi_{23i}}{y_j} \d V = 0 \comma 
\int_\Omega  C\mi_{ijkl}  L_{13kl}  \pd{\del \varphi_{13i}}{y_j} \d V = 0 \comma 
\int_\Omega  C\mi_{ijkl}  L_{12kl}  \pd{\del \varphi_{12i}}{y_j} \d V = 0 \comma 
\eqend
in order to obtain $\varphi_{11i}$, $\varphi_{22i}$, $\varphi_{33i}$, $\varphi_{23i}$, $\varphi_{13i}$, $\varphi_{12i}$, respectively. We use these values in Eq.\,\eqref{solve2}. This method is admissible under the assumption that for each $ab$ in \textsc{Voigt}'s notation indices, $\t\varphi$ components are \textit{per se} independent. Also the use in Eq.\,\eqref{solve2} is justified since we obtain $21$ components of the stiffness tensor as follows:
\begeq
C\ma_{1111} = \frac1V \int_\Omega C\mi_{ijkl} L_{11ij} L_{11kl} \d V \comma L_{11kl} = \delta_{1k}\delta_{1l} + \pd{\varphi_{11k}}{y_l} \ , \\
C\ma_{1122} = \frac1V \int_\Omega C\mi_{ijkl} L_{11ij} L_{22kl} \d V \comma L_{22kl} = \delta_{2k}\delta_{2l} + \pd{\varphi_{22k}}{y_l} \ , \\
\dots \\
C\ma_{1212} = \frac1V \int_\Omega C\mi_{ijkl} L_{12ij} L_{12kl} \d V \comma L_{12kl} = \delta_{1k}\delta_{2l} + \pd{\varphi_{12k}}{y_l} \ .
\eqend
Of course, depending on the substructure, it may be the case that some of $\t\varphi$ components are equivalent; however, this symmetry is metamaterial specific. In the same manner, from Eq.\,\eqref{solve6}, we use $\psi_{abci}=\psi_{baci}$ and for $i=1$ we solve
\begeq
\int_\Omega \bigg( -C\ma_{1c11} \del\psi_{11c1} 
+ C\mi_{1ckl} L_{11kl} \del\psi_{11c1} 
-  \Big(
C\mi_{1jkl} N_{11ckl}
\Big) \pd{\del\psi_{11c1} }{y_j} \bigg) \d V
= 0 \ , \\
\int_\Omega \bigg( -C\ma_{1c22} \del\psi_{22c1} 
+ C\mi_{1ckl} L_{22kl} \del\psi_{22c1} 
-  \Big(
C\mi_{1jkl} N_{22ckl}
\Big) \pd{\del\psi_{22c1} }{y_j} \bigg) \d V
= 0 \ , \\
\int_\Omega \bigg( -C\ma_{1c33} \del\psi_{33c1} 
+ C\mi_{1ckl} L_{33kl} \del\psi_{33c1} 
-  \Big(
C\mi_{1jkl} N_{33ckl}
\Big) \pd{\del\psi_{33c1} }{y_j} \bigg) \d V
= 0 \ , \\
\int_\Omega \bigg( -C\ma_{1c23} \del\psi_{23c1} 
+ C\mi_{1ckl} L_{23kl} \del\psi_{23c1} 
-  \Big(
C\mi_{1jkl} N_{23ckl}
\Big) \pd{\del\psi_{23c1} }{y_j} \bigg) \d V
= 0 \ , \\
\int_\Omega \bigg( -C\ma_{1c13} \del\psi_{13c1} 
+ C\mi_{1ckl} L_{13kl} \del\psi_{13c1} 
-  \Big(
C\mi_{1jkl} N_{13ckl}
\Big) \pd{\del\psi_{13c1} }{y_j} \bigg) \d V
= 0 \ , \\
\int_\Omega \bigg( -C\ma_{1c12} \del\psi_{12c1} 
+ C\mi_{1ckl} L_{12kl} \del\psi_{12c1} 
-  \Big(
C\mi_{1jkl} N_{12ckl}
\Big) \pd{\del\psi_{12c1} }{y_j} \bigg) \d V
= 0 \ , 
\eqend
for $i=2$
\begeq
\int_\Omega \bigg( -C\ma_{2c11} \del\psi_{11c2} 
+ C\mi_{2ckl} L_{11kl} \del\psi_{11c2} 
-  \Big(
C\mi_{2jkl} N_{11ckl}
\Big) \pd{\del\psi_{11c2} }{y_j} \bigg) \d V
= 0 \ , \\
\dots \\
\int_\Omega \bigg( -C\ma_{2c12} \del\psi_{12c2} 
+ C\mi_{2ckl} L_{12kl} \del\psi_{12c2} 
-  \Big(
C\mi_{2jkl} N_{12ckl}
\Big) \pd{\del\psi_{12c2} }{y_j} \bigg) \d V
= 0 \ , 
\eqend
for $i=3$
\begeq
\int_\Omega \bigg( -C\ma_{3c11} \del\psi_{11c3} 
+ C\mi_{3ckl} L_{11kl} \del\psi_{11c3} 
-  \Big(
C\mi_{3jkl} N_{11ckl}
\Big) \pd{\del\psi_{11c3} }{y_j} \bigg) \d V
= 0 \ , \\
\dots \\
\int_\Omega \bigg( -C\ma_{3c12} \del\psi_{12c3} 
+ C\mi_{3ckl} L_{12kl} \del\psi_{12c3} 
-  \Big(
C\mi_{3jkl} N_{12ckl}
\Big) \pd{\del\psi_{12c3} }{y_j} \bigg) \d V
= 0 \ .
\eqend
In this way, we solve for $\psi_{11c1}\dots \psi_{12c3}$ separately and use them to obtain $\t G\ma$ and $\t D\ma$ by means of Eq.\,\eqref{solve5}.

\section{Results and discussion}

\begin{figure}
	\centering
\includegraphics[width=0.9\textwidth]{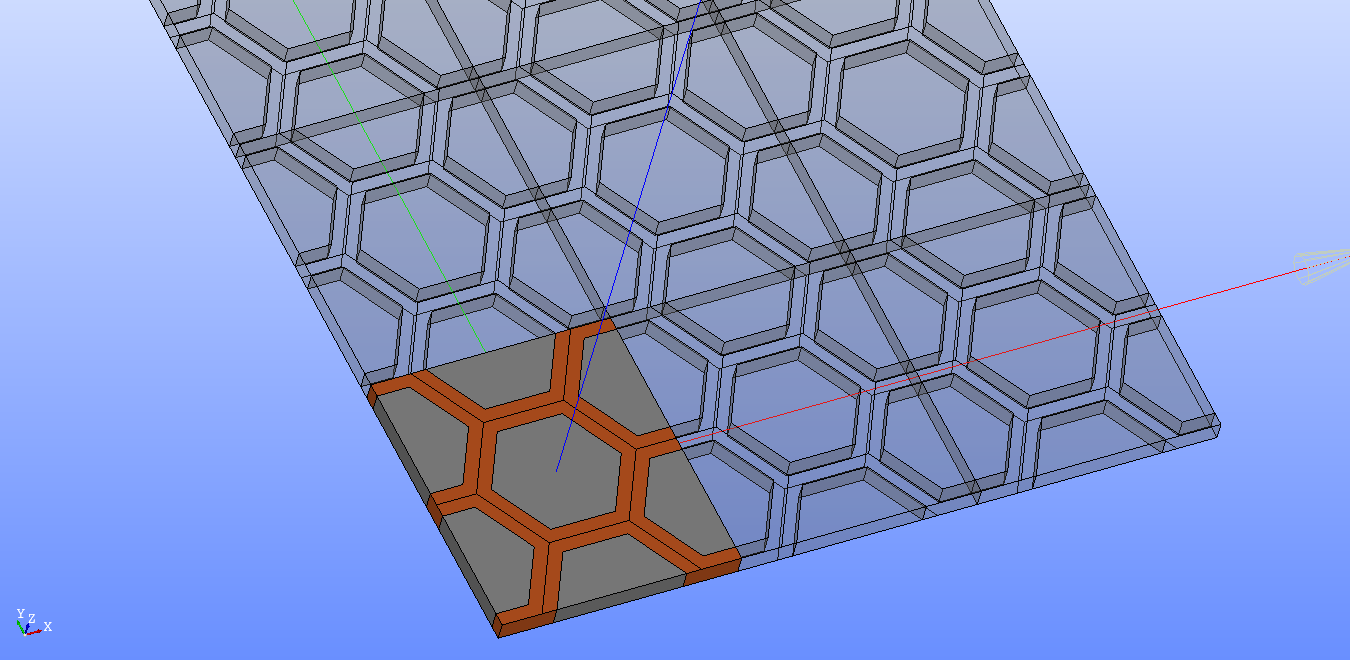}
	\caption{Honeycomb structure in Salome and a possible representative volume element (RVE) shown opaque within the transparent structure, orange denotes the 3-D printed material and gray is void (air) modeled with a significantly low modulus}
	\label{fig:geometry}
\end{figure}

By virtue of 3-D printers, it is possible to manufacture complex structures with voids inside. Voids result in a porous structure at the microscale. We stress that the voids are introduced on purpose and we assume that the microscale material is full. For example in Fused Deposition Modeling (FDM), the filaments are made of non-porous material and the porosity is caused by design. This layer-by-layer manufacturing technique is coded by a software called slicer. Slicer converts the structure from the CAD design into a G-code providing the motion of the nozzle laying the melt material, i.e. print the material as a thick viscous fluid located at the given positions. For the purpose of weight reduction, all slicer softwares introduce an infill ratio, exchanging the full material with a pre-configured periodic lattice structure. Decreasing the infill ratio increases the porosity at the macroscale. One such typical honeycomb structure is a hexagonal lattice configuration as seen in Fig.\ref{fig:geometry}, the CAD is utilized in Salome, the open-source integration platform for numerical simulation.
The full material is replaced with this configuration, for which we compute the higher order terms for any homothetic ratio, $\epsilon$, with the assumption that the linear isotropic material at the microscale might be linear anisotropic strain gradient at the macroscale. For the particular RVE as seen in Fig.\ref{fig:geometry}, the homothetic ratio is unity, i.e. the infill ratio is around 50\% meaning that the half of the space is filled with the (orange) material. The homothetic ratio is inversely related to the infill ratio, for decreasing $\epsilon$ the infill ratio increases, where $\epsilon=0$ reads 100\% infill ratio meaning that the material is full and no substructure emerges. Obviously, for 100\% infill ratio, the higher order terms, $\t G\ma$, $\t D\ma$ vanish in Eq.\,\eqref{solve5}.

\begin{figure}
	\centering
\includegraphics[width=0.8\textwidth]{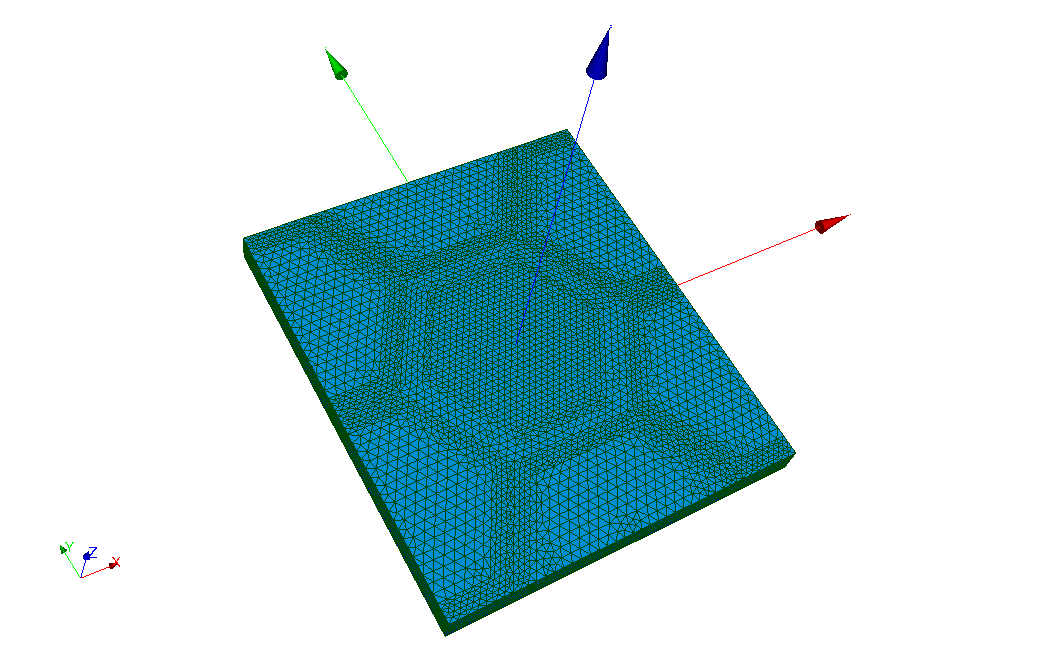}
	\caption{Used mesh of $68\,371$ tetrahedrons for the RVE, leading to $15\,618$ nodes, triangulation is obtained in Salome by using NetGen and Mephisto algorithms}
	\label{fig:mesh}
\end{figure}

\begin{figure}
	\centering
\includegraphics[width=0.7\textwidth]{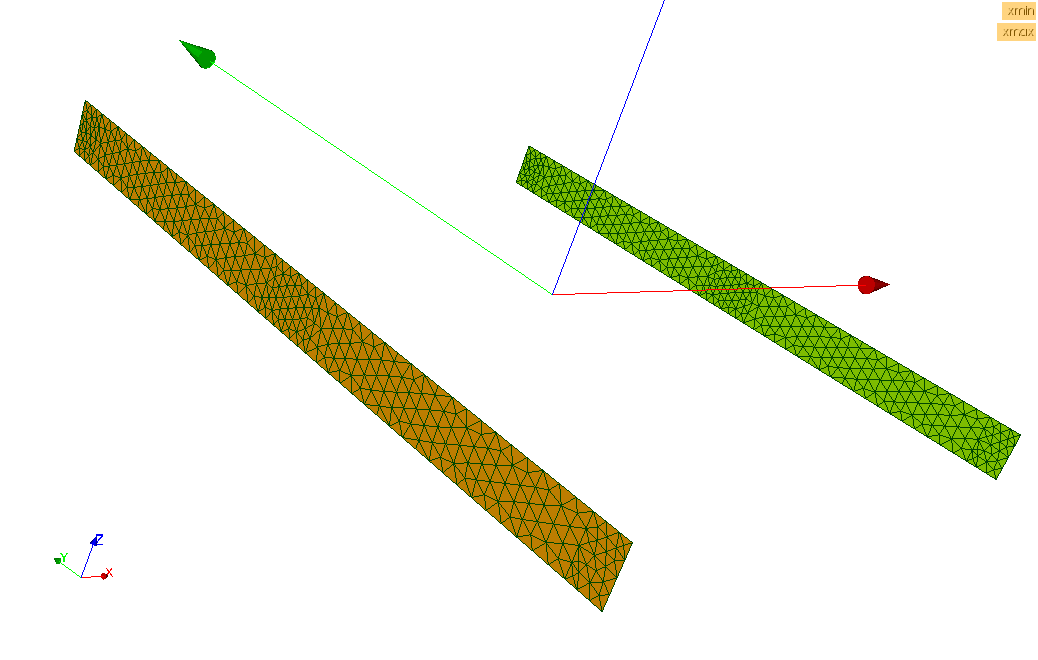}
	\caption{Demonstration of the periodic boundary conditions along $X$-axis, the same mesh is used such that the $Y$ and $Z$ coordinates are matching for nodes to be defined as the same degree of freedom}
	\label{fig:mesh2}
\end{figure}

By using the RVE, the mesh is generated in Salome by using NetGen and Mephisto algorithms as seen in Fig.\,\ref{fig:mesh}. 
We emphasis that the periodic boundary conditions need corresponding meshes on the ``neighboring'' surfaces. An example is demonstrated in Fig.\,\ref{fig:mesh2}, where along the $X_1=X$ axis, the boundary surfaces are visible. 
All nodes on both surfaces have the same $X_2=Y$ and $X_3=Z$ coordinates such that the degrees of freedom on each node are set equivalent to the corresponding node on the neighboring surface. As the periodic boundaries reflect the given solution, they are \textsc{Dirichlet} boundary conditions, which means that the macroscale and microscale solutions match along the boundaries as well. Although this condition is not \textit{a priori} set into the formulation, the use of RVE enforces matching boundaries. From the computational point of view, using \textsc{Dirichlet} boundary conditions on all surfaces, makes the problem well-defined. Hence, there are no emerging numerical problems, where we used multifrontal massively parallel sparse direct solver (mumps) for solving the weak forms and \textsc{Gauss}ian quadrature for integration.

As usual, we write out the stiffness tensor in \textsc{Voigt}'s notation with $A$, $B$ standing for combination of two indices in the order: $11$, $22$, $33$, $23$, $13$, $12$ such that the rank four tensor, $C\ma_{ijkl}$, is represented in a matrix notation,
\begeq
C\ma_{AB} = \begin{pmatrix} 
C\ma_{1111} & C\ma_{1122} & C\ma_{1133} & C\ma_{1123} & C\ma_{1113} & C\ma_{1112} \\
C\ma_{2211} & C\ma_{2222} & C\ma_{2233} & C\ma_{2223} & C\ma_{2213} & C\ma_{2212} \\
C\ma_{3311} & C\ma_{3322} & C\ma_{3333} & C\ma_{3323} & C\ma_{3313} & C\ma_{3312} \\
C\ma_{2311} & C\ma_{2322} & C\ma_{2333} & C\ma_{2323} & C\ma_{2313} & C\ma_{2312} \\
C\ma_{1311} & C\ma_{1322} & C\ma_{1333} & C\ma_{1323} & C\ma_{1313} & C\ma_{1312} \\
C\ma_{1211} & C\ma_{1222} & C\ma_{1233} & C\ma_{1223} & C\ma_{1213} & C\ma_{1212} \\    
\end{pmatrix} \ ,
\eqend
where obviously the major symmetry holds true, $C\ma_{AB}=C\ma_{BA}$, although this identity is not explicitly stated in the notation. Analogously we use $\alpha$, $\beta$ for three indices in the order: $111$, $221$, $331$, $231$, $131$, $121$, $112$, $222$, $332$, $232$, $132$, $122$, $113$, $223$, $333$, $233$, $133$, $123$ in order to be able to represent higher order terms in a matrix form as well. Specifically, for $G\ma_{ijklm}$ we have
\begeq
G\ma_{A\alpha} =
\resizebox{.85\textwidth}{!}{$\displaystyle
\begin{pmatrix}
G\ma_{11111} & G\ma_{11221} & G\ma_{11331} & G\ma_{11231} & G\ma_{11131} & G\ma_{11121} & 
G\ma_{11112} & G\ma_{11222} & G\ma_{11332} & G\ma_{11232} & G\ma_{11132} & G\ma_{11122} & 
G\ma_{11113} & G\ma_{11223} & G\ma_{11333} & G\ma_{11233} & G\ma_{11133} & G\ma_{11123} 
\\ 
G\ma_{22111} & G\ma_{22221} & G\ma_{22331} & G\ma_{22231} & G\ma_{22131} & G\ma_{22121} & 
G\ma_{22112} & G\ma_{22222} & G\ma_{22332} & G\ma_{22232} & G\ma_{22132} & G\ma_{22122} & 
G\ma_{22113} & G\ma_{22223} & G\ma_{22333} & G\ma_{22233} & G\ma_{22133} & G\ma_{22123} 
\\ 
G\ma_{33111} & G\ma_{33221} & G\ma_{33331} & G\ma_{33231} & G\ma_{33131} & G\ma_{33121} & 
G\ma_{33112} & G\ma_{33222} & G\ma_{33332} & G\ma_{33232} & G\ma_{33132} & G\ma_{33122} & 
G\ma_{33113} & G\ma_{33223} & G\ma_{33333} & G\ma_{33233} & G\ma_{33133} & G\ma_{33123} 
\\ 
G\ma_{23111} & G\ma_{23221} & G\ma_{23331} & G\ma_{23231} & G\ma_{23131} & G\ma_{23121} & 
G\ma_{23112} & G\ma_{23222} & G\ma_{23332} & G\ma_{23232} & G\ma_{23132} & G\ma_{23122} & 
G\ma_{23113} & G\ma_{23223} & G\ma_{23333} & G\ma_{23233} & G\ma_{23133} & G\ma_{23123} 
\\ 
G\ma_{13111} & G\ma_{13221} & G\ma_{13331} & G\ma_{13231} & G\ma_{13131} & G\ma_{13121} & 
G\ma_{13112} & G\ma_{13222} & G\ma_{13332} & G\ma_{13232} & G\ma_{13132} & G\ma_{13122} & 
G\ma_{13113} & G\ma_{13223} & G\ma_{13333} & G\ma_{13233} & G\ma_{13133} & G\ma_{13123} 
\\ 
G\ma_{12111} & G\ma_{12221} & G\ma_{12331} & G\ma_{12231} & G\ma_{12131} & G\ma_{12121} & 
G\ma_{12112} & G\ma_{12222} & G\ma_{12332} & G\ma_{12232} & G\ma_{12132} & G\ma_{12122} & 
G\ma_{12113} & G\ma_{12223} & G\ma_{12333} & G\ma_{12233} & G\ma_{12133} & G\ma_{12123} 
\\ 
\end{pmatrix}
$}
\ ,
\eqend
and for $D\ma_{ijklmn}$ we obtain
\begeq
D\ma_{\alpha\beta} =
\resizebox{.85\textwidth}{!}{$\displaystyle
\begin{pmatrix}
D\ma_{111 111} & D\ma_{111 221} & D\ma_{111 331} & D\ma_{111 231} & D\ma_{111 131} & D\ma_{111 121} & 
D\ma_{111 112} & D\ma_{111 222} & D\ma_{111 332} & D\ma_{111 232} & D\ma_{111 132} & D\ma_{111 122} & 
D\ma_{111 113} & D\ma_{111 223} & D\ma_{111 333} & D\ma_{111 233} & D\ma_{111 133} & D\ma_{111 123} \\ 
D\ma_{221 111} & D\ma_{221 221} & D\ma_{221 331} & D\ma_{221 231} & D\ma_{221 131} & D\ma_{221 121} & 
D\ma_{221 112} & D\ma_{221 222} & D\ma_{221 332} & D\ma_{221 232} & D\ma_{221 132} & D\ma_{221 122} & 
D\ma_{221 113} & D\ma_{221 223} & D\ma_{221 333} & D\ma_{221 233} & D\ma_{221 133} & D\ma_{221 123} \\ 
D\ma_{331 111} & D\ma_{331 221} & D\ma_{331 331} & D\ma_{331 231} & D\ma_{331 131} & D\ma_{331 121} & 
D\ma_{331 112} & D\ma_{331 222} & D\ma_{331 332} & D\ma_{331 232} & D\ma_{331 132} & D\ma_{331 122} & 
D\ma_{331 113} & D\ma_{331 223} & D\ma_{331 333} & D\ma_{331 233} & D\ma_{331 133} & D\ma_{331 123} \\ 
D\ma_{231 111} & D\ma_{231 221} & D\ma_{231 331} & D\ma_{231 231} & D\ma_{231 131} & D\ma_{231 121} & 
D\ma_{231 112} & D\ma_{231 222} & D\ma_{231 332} & D\ma_{231 232} & D\ma_{231 132} & D\ma_{231 122} & 
D\ma_{231 113} & D\ma_{231 223} & D\ma_{231 333} & D\ma_{231 233} & D\ma_{231 133} & D\ma_{231 123} \\ 
D\ma_{131 111} & D\ma_{131 221} & D\ma_{131 331} & D\ma_{131 231} & D\ma_{131 131} & D\ma_{131 121} & 
D\ma_{131 112} & D\ma_{131 222} & D\ma_{131 332} & D\ma_{131 232} & D\ma_{131 132} & D\ma_{131 122} & 
D\ma_{131 113} & D\ma_{131 223} & D\ma_{131 333} & D\ma_{131 233} & D\ma_{131 133} & D\ma_{131 123} \\
D\ma_{121 111} & D\ma_{121 221} & D\ma_{121 331} & D\ma_{121 231} & D\ma_{121 131} & D\ma_{121 121} & 
D\ma_{121 112} & D\ma_{121 222} & D\ma_{121 332} & D\ma_{121 232} & D\ma_{121 132} & D\ma_{121 122} & 
D\ma_{121 113} & D\ma_{121 223} & D\ma_{121 333} & D\ma_{121 233} & D\ma_{121 133} & D\ma_{121 123} 
\\ 
D\ma_{112 111} & D\ma_{112 221} & D\ma_{112 331} & D\ma_{112 231} & D\ma_{112 131} & D\ma_{112 121} & 
D\ma_{112 112} & D\ma_{112 222} & D\ma_{112 332} & D\ma_{112 232} & D\ma_{112 132} & D\ma_{112 122} & 
D\ma_{112 113} & D\ma_{112 223} & D\ma_{112 333} & D\ma_{112 233} & D\ma_{112 133} & D\ma_{112 123} \\ 
D\ma_{222 111} & D\ma_{222 221} & D\ma_{222 331} & D\ma_{222 231} & D\ma_{222 131} & D\ma_{222 121} & 
D\ma_{222 112} & D\ma_{222 222} & D\ma_{222 332} & D\ma_{222 232} & D\ma_{222 132} & D\ma_{222 122} & 
D\ma_{222 113} & D\ma_{222 223} & D\ma_{222 333} & D\ma_{222 233} & D\ma_{222 133} & D\ma_{222 123} \\ 
D\ma_{332 111} & D\ma_{332 221} & D\ma_{332 331} & D\ma_{332 231} & D\ma_{332 131} & D\ma_{332 121} & 
D\ma_{332 112} & D\ma_{332 222} & D\ma_{332 332} & D\ma_{332 232} & D\ma_{332 132} & D\ma_{332 122} & 
D\ma_{332 113} & D\ma_{332 223} & D\ma_{332 333} & D\ma_{332 233} & D\ma_{332 133} & D\ma_{332 123} \\ 
D\ma_{232 111} & D\ma_{232 221} & D\ma_{232 331} & D\ma_{232 231} & D\ma_{232 131} & D\ma_{232 121} & 
D\ma_{232 112} & D\ma_{232 222} & D\ma_{232 332} & D\ma_{232 232} & D\ma_{232 132} & D\ma_{232 122} & 
D\ma_{232 113} & D\ma_{232 223} & D\ma_{232 333} & D\ma_{232 233} & D\ma_{232 133} & D\ma_{232 123} \\ 
D\ma_{132 111} & D\ma_{132 221} & D\ma_{132 331} & D\ma_{132 231} & D\ma_{132 131} & D\ma_{132 121} & 
D\ma_{132 112} & D\ma_{132 222} & D\ma_{132 332} & D\ma_{132 232} & D\ma_{132 132} & D\ma_{132 122} & 
D\ma_{132 113} & D\ma_{132 223} & D\ma_{132 333} & D\ma_{132 233} & D\ma_{132 133} & D\ma_{132 123} \\
D\ma_{122 111} & D\ma_{122 221} & D\ma_{122 331} & D\ma_{122 231} & D\ma_{122 131} & D\ma_{122 121} & 
D\ma_{122 112} & D\ma_{122 222} & D\ma_{122 332} & D\ma_{122 232} & D\ma_{122 132} & D\ma_{122 122} & 
D\ma_{122 113} & D\ma_{122 223} & D\ma_{122 333} & D\ma_{122 233} & D\ma_{122 133} & D\ma_{122 123} 
\\
D\ma_{113 111} & D\ma_{113 221} & D\ma_{113 331} & D\ma_{113 231} & D\ma_{113 131} & D\ma_{113 121} & 
D\ma_{113 112} & D\ma_{113 222} & D\ma_{113 332} & D\ma_{113 232} & D\ma_{113 132} & D\ma_{113 122} & 
D\ma_{113 113} & D\ma_{113 223} & D\ma_{113 333} & D\ma_{113 233} & D\ma_{113 133} & D\ma_{113 123} \\ 
D\ma_{223 111} & D\ma_{223 221} & D\ma_{223 331} & D\ma_{223 231} & D\ma_{223 131} & D\ma_{223 121} & 
D\ma_{223 112} & D\ma_{223 222} & D\ma_{223 332} & D\ma_{223 232} & D\ma_{223 132} & D\ma_{223 122} & 
D\ma_{223 113} & D\ma_{223 223} & D\ma_{223 333} & D\ma_{223 233} & D\ma_{223 133} & D\ma_{223 123} \\ 
D\ma_{333 111} & D\ma_{333 221} & D\ma_{333 331} & D\ma_{333 231} & D\ma_{333 131} & D\ma_{333 121} & 
D\ma_{333 112} & D\ma_{333 222} & D\ma_{333 332} & D\ma_{333 232} & D\ma_{333 132} & D\ma_{333 122} & 
D\ma_{333 113} & D\ma_{333 223} & D\ma_{333 333} & D\ma_{333 233} & D\ma_{333 133} & D\ma_{333 123} \\ 
D\ma_{233 111} & D\ma_{233 221} & D\ma_{233 331} & D\ma_{233 231} & D\ma_{233 131} & D\ma_{233 121} & 
D\ma_{233 112} & D\ma_{233 222} & D\ma_{233 332} & D\ma_{233 232} & D\ma_{233 132} & D\ma_{233 122} & 
D\ma_{233 113} & D\ma_{233 223} & D\ma_{233 333} & D\ma_{233 233} & D\ma_{233 133} & D\ma_{233 123} \\ 
D\ma_{133 111} & D\ma_{133 221} & D\ma_{133 331} & D\ma_{133 231} & D\ma_{133 131} & D\ma_{133 121} & 
D\ma_{133 112} & D\ma_{133 222} & D\ma_{133 332} & D\ma_{133 232} & D\ma_{133 132} & D\ma_{133 122} & 
D\ma_{133 113} & D\ma_{133 223} & D\ma_{133 333} & D\ma_{133 233} & D\ma_{133 133} & D\ma_{133 123} \\
D\ma_{123 111} & D\ma_{123 221} & D\ma_{123 331} & D\ma_{123 231} & D\ma_{123 131} & D\ma_{123 121} & 
D\ma_{123 112} & D\ma_{123 222} & D\ma_{123 332} & D\ma_{123 232} & D\ma_{123 132} & D\ma_{123 122} & 
D\ma_{123 113} & D\ma_{123 223} & D\ma_{123 333} & D\ma_{123 233} & D\ma_{123 133} & D\ma_{123 123} 
\\
\end{pmatrix}
$}
\ ,
\eqend
where the symmetry holds true, $D\ma_{\alpha\beta}=D\ma_{\alpha\beta}$. Therefore, we determine $21$ components for $C\ma_{AB}$, $108$ components for $G\ma_{A\alpha}$, and $171$ components for $D\ma_{\alpha\beta}$ in this work for the honeycomb structure by means of the approach explained in Eqs.\,\eqref{solve1}-\eqref{solve5}. 

Computed for an RVE of $240$\,mm\,$\times\,277.12$\,mm\,$\times\,20$\,mm along $X$, $Y$, $Z$ axes, respectively, made of an isotropic material with the \textsc{Young}'s modulus of $110$\,GPa and \textsc{Poisson}'s ratio of $0.35$, we demonstrate the results in \textsc{Voigt}-like notation introduced above. For the stiffness tensor, we obtain
\begeq
C\ma_{AB} = \begin{pmatrix} 
16 & 10 & 9 & 0 & 0 & 0 \\
10 & 11 & 7 & 0 & 0 & 0 \\
9 & 7 & 43 & 0 & 0 & 0 \\
0 & 0 & 0 & 8 & 0 & 0 \\
0 & 0 & 0 & 0 & 8 & 0 \\
0 & 0 & 0 & 0 & 0 & 3 \\    
\end{pmatrix} \text{\,GPa} \ ,
\eqend
where we round off $0.1$\,GPa in all components. For the higher order terms, results depend on the arbitrary infill ratio set by the homothetic ratio $\epsilon$, as follows:
\begeq
G\ma_{A\alpha} = \epsilon
\resizebox{.85\textwidth}{!}{$\displaystyle
\begin{pmatrix}
70 & -85 & -7 & 1 & 8 & 55 & 
-21 & 84 & 26 & -18 & -3 & -40 & 
4 & -38 & -18 & 7 & 11 & 9 
\\ 
44 & -51 & -4 & 0 & 5 & 34 & 
-24 & 96 & 30 & -21 & -4 & -46 & 
3 & -31 & -15 & 6 & 9 & 7 
\\ 
40 & -48 & -4 & 0 & 4 & 31 & 
-16 & 63 & 20 & -14 & -2 & -30 & 
19 & -178 & -83 & 35 & 51 & 41
\\ 
0 & 0 & 0 & 0 & 0 & 0 & 
3 & -31 & -15 & 6 & 9 & 7 & 
-14 & 65 & 21 & -14 & -2 & -31 
\\ 
4 & -34 & -16 & 7 & 10 & 8 & 
0 & 0 & 0 & 0 & 0 & 0 & 
37 & -44 & -4 & 0 & 4 & 28 
\\ 
-5 & 23 & 7 & -4 & 0 & -11 & 
11 & -14 & -1 & 0 & 1 & 10 & 
0 & 0 & 0 & 0 & 0 & 0
\\ 
\end{pmatrix}
$} \text{\,kN/mm}
\ ,
\eqend
with $\pm 0.1$\,kN/mm accuracy as well as 
\begeq
D\ma_{\alpha\beta} = \epsilon^2
\resizebox{.85\textwidth}{!}{$\displaystyle
\begin{pmatrix}
-102 & -63 & -58 & 0 & 0 & 0 & 
0 & 0 & 0 & 0 & 0 & 0 & 
0 & 0 & 0 & 0 & 0 & 0 \\ 
-63 & -72 & -47 & 0 & 0 & 0 & 
0 & 0 & 0 & 0 & 0 & 0 & 
0 & 0 & 0 & 0 & 0 & 0 \\ 
-58 & -47 & -275 & 0 & 0 & 0 & 
0 & 0 & 0 & 0 & 0 & 0 & 
0 & 0 & 0 & 0 & 0 & 0 \\ 
0 & 0 & 0 & -48 & 0 & 0 & 
0 & 0 & 0 & 0 & 0 & 0 & 
0 & 0 & 0 & 0 & 0 & 0 \\ 
0 & 0 & 0 & 0 & -53 & 0 & 
0 & 0 & 0 & 0 & 0 & 0 & 
0 & 0 & 0 & 0 & 0 & 0 \\ 
0 & 0 & 0 & 0 & 0 & -16 & 
0 & 0 & 0 & 0 & 0 & 0 & 
0 & 0 & 0 & 0 & 0 & 0 
\\ 
0 & 0 & 0 & 0 & 0 & 0 &
-136 & -84 & -77 & 0 & 0 & 0 & 
0 & 0 & 0 & 0 & 0 & 0 \\ 
0 & 0 & 0 & 0 & 0 & 0 &
-84 & -96 & -63 & 0 & 0 & 0 & 
0 & 0 & 0 & 0 & 0 & 0 \\ 
0 & 0 & 0 & 0 & 0 & 0 &
-77 & -63 & -366 & 0 & 0 & 0 & 
0 & 0 & 0 & 0 & 0 & 0 \\ 
0 & 0 & 0 & 0 & 0 & 0 &
0 & 0 & 0 & -64 & 0 & 0 & 
0 & 0 & 0 & 0 & 0 & 0 \\ 
0 & 0 & 0 & 0 & 0 & 0 &
0 & 0 & 0 & 0 & -70 & 0 &
0 & 0 & 0 & 0 & 0 & 0 \\
0 & 0 & 0 & 0 & 0 & 0 &
0 & 0 & 0 & 0 & 0 & -22 &
0 & 0 & 0 & 0 & 0 & 0
\\
0 & 0 & 0 & 0 & 0 & 0 &
0 & 0 & 0 & 0 & 0 & 0 &
0 & 0 & 0 & 0 & 0 & 0 \\ 
0 & 0 & 0 & 0 & 0 & 0 &
0 & 0 & 0 & 0 & 0 & 0 &
0 & 0 & 0 & 0 & 0 & 0 \\ 
0 & 0 & 0 & 0 & 0 & 0 &
0 & 0 & 0 & 0 & 0 & 0 &
0 & 0 & 0 & 0 & 0 & 0 \\ 
0 & 0 & 0 & 0 & 0 & 0 &
0 & 0 & 0 & 0 & 0 & 0 &
0 & 0 & 0 & 0 & 0 & 0 \\ 
0 & 0 & 0 & 0 & 0 & 0 &
0 & 0 & 0 & 0 & 0 & 0 &
0 & 0 & 0 & 0 & 0 & 0 \\ 
0 & 0 & 0 & 0 & 0 & 0 &
0 & 0 & 0 & 0 & 0 & 0 &
0 & 0 & 0 & 0 & 0 & 0 
\\
\end{pmatrix}
$} \text{\,TN}
\ ,
\eqend
with $0.1$\,TN accuracy, where $1$\,TN$\hat =10^{12}$\,N. A general sensitivity analysis of higher order terms is inadequate, in other words, comparison between the displacement altering because of $\t G\ma$ and $\t D\ma$ components is impossible. The structure dependence on the homothetic ratio $\epsilon$ as well as loading and boundary conditions affect the sensitivity. Therefore, we have written out all terms with their own accuracy and circumvent ourselves from reducing the complexity of the outcome. 

Since the topology is hexagonal, centro-symmetry is lacking such that $\t G\ma$ tensor of rank 5 fails to vanish. All components $D\ma_{\times33\times\times\times}$ regarding the second gradient along $Z$-axis are zero due to the chosen geometry. Obviously, the periodic boundaries along $Z$-axis create hollow hexagonal tubes without ``porosity.'' Such a porous structure is indeed the case in $XY$-plane. Therefore, out of $XY$-plane the homogenization introduces a weakened structure, visible as $C\ma_{3333}$ being less than the half of the \textsc{Young}'s modulus of the material itself; however, no higher order terms occur. 

It is challenging to directly relate the homothetic ratio to the physical length scale and further studies are necessary in order to justify this study's parameters.

\section{Conclusions}

Generalized mechanics has been already studied in 1950s as a purely academic research. Additive manufacturing opens the door for crafting structures with substructures (microscale), called infills, leading to different length scales performing simultaneously at the macroscale, thus, making the generalized elasticity necessary for accurate modeling. Involving strains, conventional elasticity necessitates 21 material parameters. Generalized elasticity with strain gradients introduces additional to the 21 (different) parameters in $\t C\ma$ rank 4 tensor, another 108 parameters in $\t G\ma$ rank 5, and 171 parameters in $\t D\ma$ rank 6 tensors. Asymptotic analysis results in micro-macro-scale relations that we briefly yet thoroughly demonstrated in this work. Finally, a new methodology is proposed for using the substructure and determining all the parameters in generalized elasticity by using computations based on the finite element method (FEM). In order to present the method on a particular case of hexagonal honeycomb substructure, open-source codes based numerical implementation is established under GNU public license \cite{gnupublic}, the code is available in \cite{compreal} in order to allow a transparent scientific exchange.

\subsection*{Acknowledgements}
B. E. Abali's work was partly funded by a grant from the Daimler and Benz Foundation.

\bibliographystyle{special}
\bibliography{homogenization}

\end{document}